\documentstyle[twocolumn,aps,prl,graphics]{revtex} 
\draft 
\title{Nucleation Induced Undulative Instability in Thin Films of $n$CB
Liquid Crystals} 
\author{Stefan Schlagowski \and Karin Jacobs \and Stephan Herminghaus} 
\address{Dept. of Applied Physics, Ulm University, D-89069 Ulm, Germany}

\begin{document}
\maketitle

\begin{abstract}
  A surface instability is reported in thin nematic films of 5CB and
  8CB, occurring near the nematic--isotropic phase transition.
  Although this instability leads to patterns reminiscent of spinodal
  dewetting, we show that it is actually based on a nucleation
  mechanism.  Its characteristic wavelength does not depend markedly
  on film thickness, but strongly on the heating rate.
\end{abstract}


Following several studies on the spreading behavior of liquid crystals
(LC) from the $n$CB homologous series (4'-$n$-alkyl-4-cyanobiphenyl)
\cite{BarS96,ValV96,BarV98,VanB98,BarO99,XuSa00}, undulative
instabilities have been observed in thin films of the LC 5AB$_4$
\cite{HerJ98}, and 5CB \cite{VanV99} near the nematic--isotropic
(N--I) phase transition. In the case of 5CB, these results have led to
some discussion whether a spinodal dewetting mechanism driven by van
der Waals forces is at work, as proposed by Vandenbrouck et al.
\cite{VanV99}, or wether the instability is driven by a
pseudo--Casimir force based on the director fluctuations in thin
nematic films \cite{AjdP91,ZihZ98,ZihP99,ZihP00,ZihK00}. In the
present paper, we show that neither is true for $n$CB thin films.
Instead, the instability is caused by textures in the nematic film
which largely determine the characteristic wavelength of the emerging
pattern.
 
5CB and 8CB were obtained from Merck KGaA (Darmstadt, Germany) and
Frinton Laboratories Inc. (Vineland, NJ) respectively, and used
without further purification. Silicon wafers (100--oriented,
p--(Boron--) doped) with a native oxide layer of 2 nm provided by
Wacker Chemitronics (Burghausen, Germany) were used as solid
substrates. The wafers were cut to samples approximately 1 cm$^2$ in
size and cleaned with a Snowjet$^{TM}$ (Tectra, Frankfurt/M, Germany),
a cold CO$_2$ stream effectively removing particulate and organic
contamination \cite{SheH94}, followed by ultrasonication in ethanol,
acetone, and hexane, subsequently.

Immediately after this cleaning process, LC films were spincast onto
the samples from hexane solutions. Variation of concentration and
spinning rate allows to deposit films of variable thickness. The
preparation procedure was performed in a class 100 clean room
environment at room temperature. Therefore, the films were initially
in the nematic (5CB) or smectic A (8CB) state, respectively.  Film
thicknesses were recorded with an ellipsometer (Optrel GbR, Berlin,
Germany). The samples were placed on a  heat stage (Linkam THMSG 600, 
temperature control better than 0.1 $^\circ$C) and observed {\it in situ}
with a Zeiss Axiophot microscope equipped with a digital camera.
Unless otherwise noted, no polarizers were used in the microscope
setup.

\begin{figure}
\begin{center}
  \includegraphics{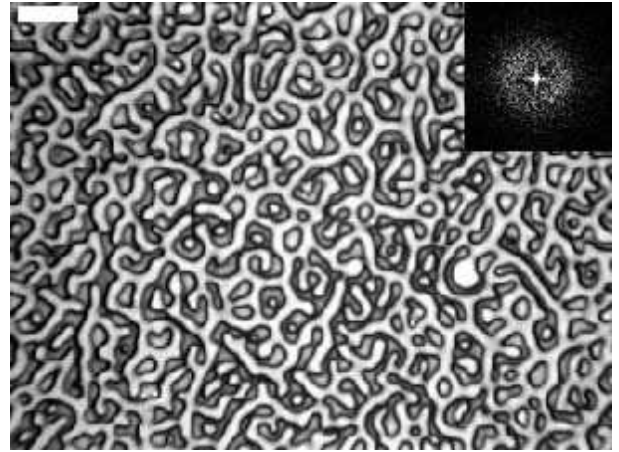}
\end{center}
\caption{\label{fig1}Surface instability of an 85.9(3) nm thick film
  of 8CB at 39.9(1) $^\circ$C. The scale bar has a length of 100 $\mu$m. The
  inset Fourier transform gives a wavelength of 40(20) $\mu$m.}
\end{figure}

Observations at room temperature showed films of 5CB and 8CB to be
stable for hours at thicknesses ranging from 50 nm to 200 nm. Upon
heating, a surface undulation with a characteristic wavelength can be
observed in both types of samples (see Figure~\ref{fig1}) close to,
but consistently below the N--I transition temperature (T$_{NI}$).

The observed undulation does not lead to a complete dewetting of the
LC film but rather disappears above T$_{NI}$, such that the film
becomes homogeneous again in the isotropic phase. Upon cooling, a
similar instability occurs. Observation under crossed polarizers
revealed that the undulation is accompanied by lateral `demixing' of
nematic and isotropic regions. 

\begin{figure}
\begin{center}
  \includegraphics{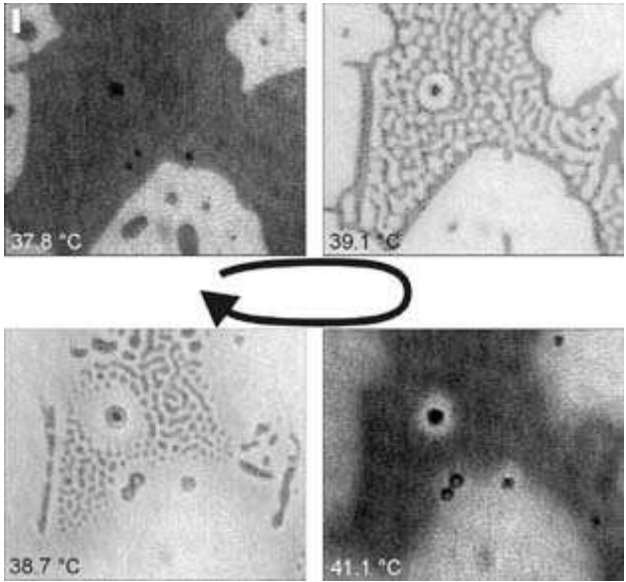}
\end{center}
\caption{\label{fig2} Heating/ cooling sequence showing the transient
  instability in a patch of 8CB (scale bar = 25 $\mu$m).}
\end{figure}

Complete dewetting of the samples occurs only by heating a few
degrees above T$_{NI}$, resulting in an array of isotropic droplets.
As heating and cooling of the sample around T$_{NI}$ was repeated, as
indicated in Fig.~\ref{fig2}, nearly identical undulation patterns
were obtained. We found that the instability developed only when the
samples were heated or cooled at a certain rate. Keeping them at a
fixed temperature close to the N--I phase transition resulted in the
nucleation of only a few holes in the film. These observations have
been made for 5CB and for 8CB, at all film thicknesses investigated.
The pattern was found to vanish faster for thicker films after
completion of the phase transition (see below).

\begin{figure}
\begin{center}
  \includegraphics{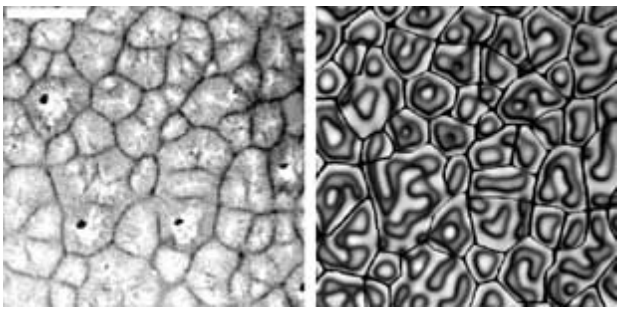}
\end{center}
\caption{\label{fig3}Left: 86 nm thick film of 8CB in the nematic
  state. Borders between nematic domains appear as dark lines. Right:
  Picture of the surface undulation superimposed on the network of
  domain borders, emphasizing the coincidence of surface undulations
  during the instability (bright areas, thick film) with domain
  borders in the nematic phase. (scale bar = 100 $\mu$m)}
\end{figure}

Careful examination of the images revealed the
formation of nematic domains in 8CB films when heated above the
smectic A--nematic transition temperature for the first time after
preparation (cf. Fig.~\ref{fig3}). This domain pattern is preserved
during subsequent heating and cooling of the samples which was
limited to a few degrees around the N--I transition temperature. On
the left hand side of the figure, contrast is enhanced in order to
clearly show the domain boundaries, while on the right hand side, the
undulative instability occurring near the phase transition is
superimposed on the domain boundary pattern obtained before on the
same spot. It is clearly seen from the overlay that the undulative
pattern is strongly correlated with the domain boundary pattern.

Since films of 5CB are nematic at room temperature, the domain pattern
observed immediately after preparation is less pronounced, but
nevertheless present.  As in the case of 8CB films, this pattern is not
influenced during repeated temperature changes of the sample.

\begin{figure}
\begin{center}
  \includegraphics{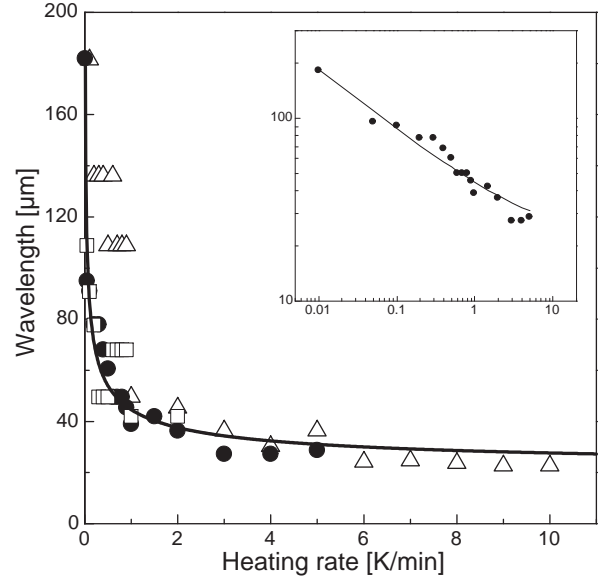}
\end{center}
\caption{\label{fig5}Dependence of undulation wavelength on heating
  rate and film thickness $h$ of 5CB, 51~nm ($\bigtriangleup$),
  87 nm ($\bullet$) and 139 nm ($\Box$). Data for $d$=87~nm are
  fitted according to Eq.~\ref{fit} and are presented seperately on a
  log--log scale in the inset.}
\end{figure}

Since we have found that the undulations appear only when the sample
temperature is swept through the phase transition, it is of interest
to investigate the impact of the heating rate. We have thus varied the
heating rate from 0.01 K/min to 10 K/min. As shown in Fig.~\ref{fig5},
the heating rate is indeed a major defining parameter for the
wavelength of the undulation: The faster the samples are heated into the
isotropic phase, the smaller is the undulation wavelength.

To explain the instability as such we propose the following scenario:
After preparation (for 5CB) or after heating to the nematic phase for
the first time (for 8CB), a pattern of nematic domains exists in the
films that is preserved during the course of the experiment, since
close to the substrate nematic or even smectic order exists even at
temperatures substantially above the clearing point (see e.g.
\cite{KOcB00}). The domain boundaries act as nucleation sites for the
isotropic phase upon further heating. It should be noted that
apparently similar observations in 5AB4 \cite{HerJ98} do not belong to
this class of nucleated undulation phenomena, since in that study the
temperature was kept constant, and the undulation was not transient,
but remained once it was formed.
  
It is interesting to note that there is no systematic dependence of
the patterns observed with $n$CB on the film thickness.  Samples of
different thickness (3 thicknesses each for 5CB and 8CB) show no
pronounced dependence of the undulation wavelength on film thickness,
as shown in Fig.~\ref{fig5}. In particular, no indication of a
quadratic dependence of the wavelength on film thickness could be
detected, which would be expected of a spindoal dewetting scenario.
 
It is possible to explain the pronounced effect of the heating rate
upon the wavelength of the undulation with a few assumptions on the
nucleation sites.  First we note that the undulations appear only in a
certain temperature window, $T_l < T < T_h$, which is more
extended and shifted to lower temperatures in thin films than in
thicker ones. For thick films $T_h$ will approach the bulk N--I phase
transition temperature in accordance to earlier studies \cite{WitL96}.
Each nucleation site present in the film (e.~g. the domain borders
seen clearly in the 8CB samples) is expected to nucleate an isotropic
domain, and concomitantly a modulation of the local film thickness,
when a certain temperature $T_s$ within the interval $\Delta T= T_h - T_l$ is
reached.This modulation, or domain then grows with a certain velocity
$v$, and thus further nucleation within the domain is precluded,
effectively reducing the number of nucleation sites to become active
at a higher temperature.  The total number density of isotropic domains $D$
that will develop in a film, and hence the dominant lateral scale of
the undulation $\lambda \approx 1/\sqrt{D}$, thus depend on the
heating rate. If $n(T_s)$ is the number density of nucleation sites
which nucleate a domain when the temperature $T_s$ is reached, $D$,
will be given by

\begin{equation}
  \frac{dD}{dt} = \alpha n(T(t)) e^{-A(t)}
\end{equation}

where $\alpha$ is the heating rate, and $A(t)$ is the total area
fraction of the domains if they are assumed circular. The exponential
takes care of the mutual overlap of these (circular) model domains.
For simplicity, assume that $n(T)=n_0=${\it const.}  for $T_l \le T
\le T_h$, and 0 otherwise. If we furthermore assume the growth rate
$dr/dt$ of the domains to be constant, we have
\begin{equation}
  \frac{d^2A}{dt^2} = 2\pi v^2 D
\end{equation}
and thus, after some rearrangements,
\begin{equation} \label{Dmaster}
  2 \pi v^2 D (D')^2 + D' D''' - (D'')^2 = 0  
\end{equation}
For investigating the effect of a temperature ramp, $T = T_l + \alpha
t$, this must be solved with the initial conditions $D(0)=0$ and
$D'(0)=\alpha n_0$.  The number density of domains formed in this way cannot
be expressed analytically, but is well approximated by

\begin{equation}\label{fit}
  D_{max} = N \frac{3 (\alpha/\alpha_0)^{2/3}}{1+3(\alpha/\alpha_0)^{2/3}}
\end{equation}

which gives the correct scaling for $\alpha \rightarrow 0$ as well as
$\alpha \rightarrow \infty$.
$N = n_0 \Delta T$ is the total number of nucleation sites and
$\alpha_0 = \sqrt{2N\pi v^2 \Delta T^2}$.

As shown in Fig.~\ref{fig5}, the fit function in Eq.~(\ref{fit}) does
reproduce the heat rate dependence of the undulation wavelength rather
well. The scatter in the calculated values for $N$ and $v$ is too
large to confirm any systematic variation with film thickness, as it
is expected for a nucleation process triggered by inhomogeneities. A
systematic dependence on film thickness, as expected for spinodal
dewetting, is not found. The different evolution of the nematic
domains in 5CB (created upon spincasting the films) and 8CB (domains
grow at first heating cycle), however, could lead to specific
differences in the two types of samples. Further measurements are
necessary to elucidate this point.

\begin{figure}
\begin{center}
  \includegraphics{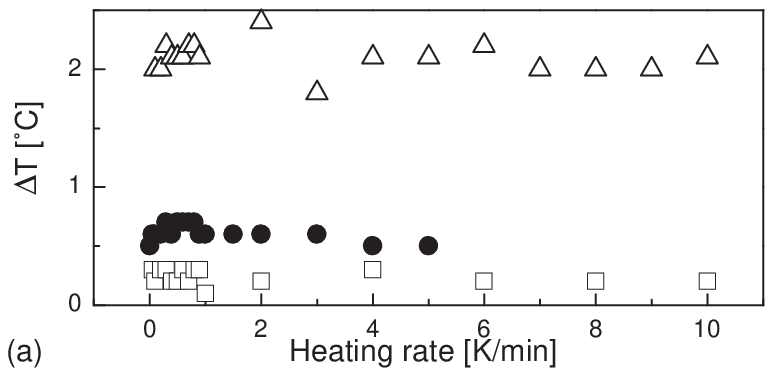} \includegraphics{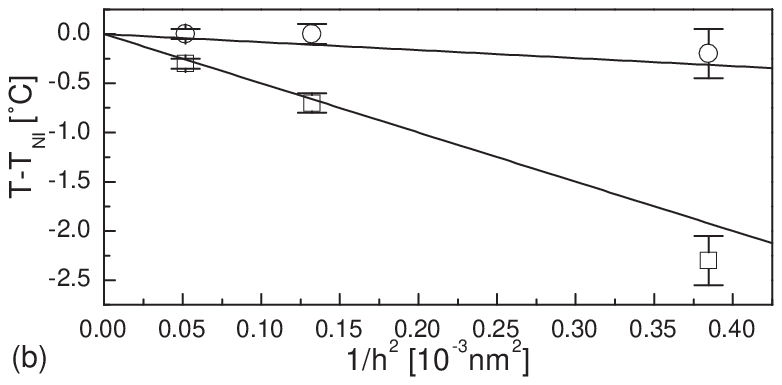}
\end{center}
\caption{\label{fig6}(a) Temperature interval $\Delta T = T_h-T_l$
  inside which the undulations are visible plotted against heating
  rate for three different film thicknesses $h$, 51~nm
  ($\bigtriangleup$), 87~nm ($\bullet$) and 139~nm ($\Box$). (b)
  $T_l-T_{NI}$ ($\bigtriangleup$) and $T_h-T_{NI}$ ($\circ$) plotted
  against $1/h^2$. $T_l$ and $T_h$ are mean values over all heating
  rates respectively. The straight lines represent data from
  \protect\cite{EffO01}. Data shown are for 5CB films, 8CB films show
  the same qualitative behavior.}
\end{figure}

It is finally of interest to investigate the temperature range $\Delta
T$, in which the modulation pattern persists. Values for $\Delta T$
(based on observation of the temperatures at which the undulation
started to appear respectively ceased to be visible) were found to be
independent of the heating rate as shown in Fig.~\ref{fig6}~(a) but
decrease with increasing film thickness, Fig.~\ref{fig6}~(b). 

The mechanism leading to different thicknesses of the nematic and
isotropic parts of the film during the instability is still not clear.
Initially, it was proposed that a Marangoni flow could drive the
instability \cite{HerF99}, since the surface tensions of liquid
crystalline compounds have been shown to exhibit a sharp increase in
the vicinity of the N--I phase transition \cite{GanF78,GeoM92,MohG92}.
Once the phase transition is completed, the gradient in surface
tension would vanish, resulting in a flat isotropic film, explaining
the transient nature of the instability. However, very recent
observations of $n$CB films kept at constant temperatures in a
temperature range near the bulk phase transition temperature have
found a nematic--isotropic phase coexistence with both phases taking a
distinct equilibrium thickness \cite{EffO01}. This effect could very
well drive the thickness modulations in our dynamic experiments.
Indeed, our plot of $\Delta T$ against $h^{-2}$ in Fig.~\ref{fig6}~(b)
is in good agreement with the data found in \cite{EffO01} by
Effenterre et.~al.  under different experimental circumstances.

To conclude, we have shown that the undulative instability observed in
thin films of $n$CB liquid crystals is not based on a spontaneous
(spinodal) dewetting mechanism, but develops during the coexistence of
the nematic and isotropic phase in the films in a temperature window
close to the N--I phase transition. Isotropic areas are nucleated at
defects (domain borders) in the nematic film and the growth and
coalescence of these areas lead to the observed undulation patterns.
Changing the heating rate during the experiments strongly influences
the undulation patterns since less isotropic areas can be nucleated at
high heating rates before the phase transition is complete. First
experiments on substrates with artificially induced nucleation sites
confirm the proposed nucleation mechanism. These findings will be
covered in a forthcoming paper.

This work was funded by the Deutsche Forschungsgemeinschaft through the
priority program 1052 `Wetting and Structure Formation at Interfaces'
under grant numbes He2016/5 and Ja905/1.  We acknowledge generous support of Si~wafers by Wacker Chemitronics, Burghausen, Germany.

\end{document}